\documentclass[aps,preprint,showpacs,preprintnumbers,amsmath,amssymb]{revtex4}
\usepackage{amsmath,mathrsfs,amsbsy,color,graphicx,bm,amsthm,amsfonts}
\usepackage{units}
\usepackage{bbm}
\usepackage{times}
\usepackage{dcolumn}
\usepackage{mathrsfs}
\usepackage{amsmath,amssymb,epsfig}

\begin{document}

\title{Parameter estimation for an expanding universe}
\author{Jieci Wang$^{1,2}$, Zehua Tian$^{1}$, Jiliang Jing$^{1}$ and Heng Fan$^{2}$ \footnote{Email: hfan@iphy.ac.cn}}
\affiliation{ $^1$ Department of Physics, and Key Laboratory of Low
Dimensional Quantum Structures and Quantum
Control of Ministry of Education,
 Hunan Normal University, Changsha, Hunan 410081, China\\
$^2$ Beijing National Laboratory for Condensed Matter Physics, Institute of Physics,Chinese Academy of Sciences, Beijing 100190, P. R. China
}


\begin{abstract}
 We study the parameter estimation for excitations of Dirac fields in the expanding Robertson-Walker universe. We employ quantum metrology techniques to demonstrate the possibility for high precision estimation for the volume rate of the expanding universe. We show that the optimal precision of the estimation depends sensitively on the dimensionless mass $\tilde{m}$ and dimensionless momentum $\tilde{k}$ of the Dirac particles. The optimal precision for the ratio estimation peaks at some finite dimensionless mass $\tilde{m}$ and momentum $\tilde{k}$. We find that the precision of the estimation can be improved by choosing the probe state as an eigenvector of the hamiltonian. This occurs because the largest quantum Fisher information is obtained by performing projective measurements implemented by the projectors onto the eigenvectors of specific probe states.
\end{abstract}

\vspace*{0.5cm}
 \pacs{03.67.-a,06.20.-f,04.62.+v }
\maketitle

\section{Introduction}
The field of quantum metrology has been developed to study the ultimate limit of precision in estimating a physical quantity, when quantum strategies are exploited \cite{advances}. Classically,
for a given measurement scheme, the effect of statistical errors can
be reduced by repeating the same measurement and averaging the outcomes.
Explicitly, by repeating the same independent measurements $N$ times,
the uncertainty about the measured parameter $\Theta$ will be reduced to
the standard quantum limit $\Delta\Theta \simeq 1/\sqrt{N}$.
However, if squeezed or entangled states are used as probe systems which undergo an evolution that depends on the parameters to be estimated, the precision can be enhanced and the uncertainty can approach the Heisenberg limit $\Delta\Theta \simeq 1/N$ which beats
the standard quantum limit \cite{Braunstein1994}.  It has been demonstrated that the sensitivity of the detectors in the Laser Interferometer Gravitational-Wave Observatory (LIGO) approaches the Heisenberg limit by using squeezed states of light \cite{Aasi}.
Furthermore, quantum entanglement plays an important role in open questions such as the information loss problem of black holes and many authors have recently recently studied how relativistic effects might affect entanglement, classical and quantum
discord as well as  quantum nonlocality \cite{RQI1,Ralph,adesso2,RQI2,RQI3,RQI4,RQI5,RQI7} in a relativistic frame. Most recently, the study of  quantum metrology has been applied to
the area of relativistic quantum information \cite{RQI6,aspachs,HoslerKok2013,downes, mostrecent,RQM,RQM2}.
It was shown that quantum metrology methods can be exploited to improve
probing technologies of Unruh-Hawking effect \cite{RQI6,aspachs,HoslerKok2013}
and gravitation \cite{downes, mostrecent}. It was found that quantum metrology can be
employed to estimate the Unruh temperature of a moving cavity at
experimental reachable acceleration~\cite{RQM,RQM2}. These preliminary results are of great
importance for the observation of relativistic effects in a laboratory \cite{casimirwilson,casimir2} and
space-based quantum information processing tasks \cite{zeilingerteleport,Bruschi1306,Bruschi1309}.

To obtain a better understanding of the application of quantum metrology to relativistic settings, it is interesting to extend the studies to curved frameworks. We notice
that the quantum metrology for scalar fields
in a relativistic frame has been the focus
of research \cite{RQI6,aspachs,HoslerKok2013,RQM,RQM2}, but such technology  has not yet been applied to
quantum parameter estimation for fermionic fields.  It is well known that neutrinos are one kind of fermions and the time-evolution of a neutrino in an expanding universe can be described by Dirac equation. In the standard cosmology model, relic neutrinos
act as witnesses and participants for landmark events in the history of the
universe from the era of big-bang nucleosynthesis to the era of large scale
structure formation \cite{Weinberg2,Steigman}. Thus, understanding the
behavior of fermions, especially the neutrinos is crucial for
estimating the age of our universe and its fate.
We also notice that the neutrinos are usually considered as
spin $1/2$ fermions while the spinless neutrinos is only discussed in few articles such as \cite{Alex}.
Here we study the massive Dirac field without the consideration of spin therefore can
not model the neutrinos directly. But the model in this paper can in principle be applied to the study of  spin $1/2$ fermion \cite{fredericivy1} in the Robertson-Walker universe as well. Besides, the estimation of the expansion of the universe is significant to understand the source and mechanisms of primary particle production.

In this paper, we apply techniques from quantum metrology to estimate the
expansion rate of the universe in the two-dimensional conformally flat  Robertson-Walker (RW) metric \cite{Birrelldavies,dun1,Bergs,fredericivy,fredericivy1,fredericivy2,fredericivy3} for Dirac fields.  The interplay between the expansion of the universe and entanglement in this context was studied in Refs. \cite{fredericivy,fredericivy1,fredericivy2,fredericivy3} both for scalar and Dirac fields. It was found that a single qubit initially in its vacuum state evolves to an entangled
state in the particle number degree of freedom in the expanding universe. We will analytically calculate the quantum
Fisher information (QFI) \cite{paris} in terms of the dimensionless mass $\tilde{m}$  and the dimensionless wave vector $\tilde{k}$ of the Dirac fields, where the classical expanding universe constitutes the unitary channel. By applying a quantum metrology strategy on the final states,
the precision of the estimated parameter can be related to the QFI. Such procedure corresponds to finding the set of measurements that allows us to estimate the parameter with the highest precision. At the same time, information about the history of the expanding universe can be extracted from measurements on the  states in the asymptotic future region, while the ultimate precision limit is related to the QFI.
We then find that the precision of the estimation can be improved by
choosing the projective measurements corresponding to eigenvectors of the particle states with some certain masses and momentums.

The outline of the paper is as follows. In Sec. II we introduce the quantization of Dirac fields in the RW spacetime. In Sec. III  some concepts for quantum metrology are presented, in particular the QFI. In Sec. IV we calculate the QFI and seek the best proposal for the cosmological parameter estimation. The last section is devoted to a brief summary.

\section{Quantization of Dirac fields in expanding universe}

In this paper we study fermions for the following two reasons. Firstly, quantum metrology strategies for scalar fields in a relativistic frame have been the focus of research in Refs. \cite{RQI6,aspachs,HoslerKok2013,RQM,RQM2}, while quantum parameter estimation for the fermionic field has not been discussed.  Secondly, the cosmic neutrino background (C$\upsilon$B) was produced approximately 1 second after the big bang and permeates the present day universe \cite{Benjamin}. It is similar to the cosmic microwave background (CMB) in the sense that both of them are relic distributions created after the big bang. However, differently from the CMB that formed after the universe was approximately four hundred thousand years old, the C$\upsilon$B formed only about one second after the big bang \cite{Weinberg2,Steigman,Benjamin}. Thus, understanding the  behavior of fermions, especially the relic neutrinos is crucial for understanding the inflationary phase of the universe and the estimating the age of our universe. For more details please see Ref. \cite{fredericivy1}.

To quantize Dirac fields in the RW spacetime \cite{Birrelldavies,dun1,Bergs}, we
introduce the local tetrad field
${\rm
g}^{\mu\nu}=e^{\mu}_{~a}(x)e^{\nu}_{~b}(x)\eta^{ab}~$,
where ${\rm g}^{\mu\nu}$ and $\eta_{ab}$ are metrics of the RW spacetime
and Minkowski spacetime, respectively. The covariant Dirac equation
for field $\Psi$ with mass $m$
in a curved background reads\cite{Brill}
\begin{equation}
\label{Diraceq} i\gamma^{\mu}(x)\left( \frac{\partial}{\partial
x^{\mu}}- \Gamma_{\mu}\right)\Psi=m\Psi~.
\end{equation}
where the curvature-dependent Dirac matrices
$\gamma^{\mu}(x)$ relate to flat ones through
$\gamma^{\mu}(x)=e^{\mu}_{~a}(x)\bar{\gamma}^{a}$,
and $\bar{\gamma}^{a}$ are the flat-space Dirac matrices that
satisfies $\{\bar{\gamma}^{a},\bar{\gamma}^{b}\}=2\eta^{ab}$. Here $\Gamma_\mu =
\frac{1}{8}[\gamma^\alpha,\gamma^\beta]e_\alpha{}^\nu e_{\beta\nu;\mu}$ are the spin
connection coefficients. Throughout this paper we adopt the conventions $\hbar=c=1$.

We solve the Dirac equation in a two-dimensional RW spacetime,
which is proved to embody all the fundamental features of
the higher dimensional counterparts \cite{fredericivy1}.
The corresponding line element reads
\begin{equation}
\label{Metric}ds^2=[a(\eta)]^2(d\eta^2-dx^2),
\end{equation}
where the conformal time $\eta$ is related to cosmological time $t$ by
$a(\eta)=\int{a^{-1}(t)dt}$. We specialize to a
conformal factor of a form (see \cite{dun1,fredericivy1,fredericivy2,fredericivy3,Bernard}), in which
the conformal
factor $a(\eta)$ can be written as
$
a(\eta)=1+\epsilon(1+\tanh\rho\eta)$, where $\epsilon$ and $\rho$ are positive real parameters, corresponding to
the ratio between the initial and final volume and the expansion rate of the universe \cite{fredericivy1,Birrelldavies}, respectively. Note that $a(\eta)$ is constant in the far past $\eta\rightarrow -\infty$ and far
future $\eta\rightarrow +\infty$, which means that the
spacetime tends to a flat Minkowskian spacetime asymptotically.
In the asymptotic past ($in$)-
and future ($out$)-regions the vacuum states
and one-particle states may be well defined \cite{Birrelldavies}. Given the particular choice of spacetime background the spin connections read
$\Gamma_{\mu}=\frac{\dot{a}(\eta)}
{4a(\eta)}[\bar{\gamma}_{\mu},\bar{\gamma}_0]$,
where $\dot{a}(\eta)=\frac{\partial a(\eta)}{\partial\eta}$.
To solve  the Dirac equation Eq. (\ref{Diraceq}) in the metric Eq.
(\ref{Metric}), we re-scale the field as
$\Psi =a^{-3/2}[\bar{\gamma}^{\nu}\partial_{\nu}-ma(\eta)]\psi$ \cite{fredericivy1,fredericivy2}, and the dynamic equation becomes
$\eta^{\mu\nu}\partial_{\mu}\partial_{\nu}\psi-
\bar{\gamma}^{0}\dot{ma(\eta)}\psi-[m a(\eta)]^2\psi=0$.
Moreover, we can represent the re-scaled
 solution $\psi$ into a spatial and
temporal separable form and find that
\begin{equation}
\ddot{\psi_k}^{(\pm)}+
\left[|{{\pm k}}|^{2}+m^2a(\eta)^2\pm
i m\dot{a}(\eta)\right]\psi_k^{(\pm)}=0.
\label{feq}
\end{equation}
Actually  $\psi_k^{(+)}$ and $\psi_k^{(-)}$ are
positive and negative  frequency modes with respect to conformal
time $\eta$ near the asymptotic past and future, i.e.
$i\dot{\psi_k}^{(\pm)}(\eta)\approx \mp \omega_{in/out}
\psi_k^{(\pm)}(\eta) $ with
$\omega_{in/out}=\sqrt{|{{\pm k}}|^2+m^2 a^2(\eta\rightarrow\mp\infty)}$
and $|{{\pm k}}|^2=k^2$.
In the asymptotic past region, the positive and negative
frequency solutions \cite{dun1,fredericivy1,dun2} of  Eq. (\ref{feq}) are
\begin{eqnarray}
\nonumber\psi^{(\pm)}_{in}&=&\exp \left[ -i\omega _{(+)}\eta-\frac{
i\omega _{(-)}}{\rho }\ln (2\cosh \rho\eta )\right]\\
&\times&\sum_{n=0}^{\infty}\frac{[\rho+i\zeta_{(-)}^{\pm}]_n
[i\zeta_{(-)}^{\mp}]_n}{[\rho-i\omega_{in}]_n} \frac{\Lambda_+^n}{2^n n!},
\end{eqnarray}
where $[q]_n=q(q+1)\ldots(q+n-1)$, $\Lambda_+=1+\tanh
 (\rho \eta )$, $\zeta_{(\pm)}^{\pm}=\omega
_{(\pm)}\pm m\epsilon$ and $\omega _{(\pm) }=(\omega _{out}\pm \omega _{in})/2$.
Similarly, one may
obtain the modes of the Dirac fields that behaving as
positive and negative frequency modes in the asymptotic future region
\begin{eqnarray}
\psi^{(\pm)}_{out}=\exp \left[ -i\omega _{(+)}\eta-\frac{
i\omega _{(-)}}{\rho }\ln (2\cosh \rho \eta )\right]
\sum_{n=0}^{\infty}\frac{[\rho+i\zeta_{(-)}^{\pm}]_n
[i\zeta_{(-)}^{\mp}]_n}{[\rho+i\omega_{out}]_n} \frac{\Lambda_-^n}{2^n n!},
\end{eqnarray}
where $\Lambda_-=1-\tanh (\rho \eta )$. The whole information about the evolution of the spacetime is encoded in $\epsilon$ and $\rho$. However, these two parameters contribute to construct one quantity, i.e., the scaling factor a($\eta$) which  affects the modes. We also notice that the expansion rate $\rho$ is a scaling quantity for timescales. Therefore, the only expanding parameter we can estimate is $\epsilon$ while $\rho$ can not be measured independently. Here we can introduce the dimensionless mass as $\tilde{m}=m/\rho$ and the dimensionless wave number as $\tilde{k}=k/\rho$.
It is now possible to quantize the field and find the
relation between the vacuum state in  the asymptotic future region and that in asymptotic past. Such relation can be described by a certain set of
 Bogoliubov transformations \cite{Birrelldavies} between $in$ and $out$ modes
$\psi_{in}^{(\pm)}(\tilde{k})={\cal A}_{\tilde{k}}^{\pm}\psi^{(\pm)}_{out}(\tilde{k})
+{\cal B}_{\tilde{k}}^{\pm}\psi^{(\mp)*}_{out}(\tilde{k})$,
where ${\cal A}^{\pm}_{\tilde{k}}$ and ${\cal B}^{\pm}_{\tilde{k}}$ are Bogoliubov coefficients that take the form
\begin{eqnarray}
{\cal A}^{\pm}_{\tilde{k}}&=&\nonumber\sqrt{\frac{\omega_{out}}{\omega_{in}}}\frac{\Gamma(1-i\omega_{in})
\Gamma(-i\omega_{out})}{\Gamma(1-i\zeta_{(+)}^{\mp})
\Gamma(-i\zeta_{(+)}^{\pm})},\\*
{\cal B}^{\pm}_{\tilde{k}}&=&\sqrt{\frac{\omega_{out}}{\omega_{in}}}
\frac{\Gamma(1-i\omega_{in})
\Gamma(i\omega_{out})}{\Gamma(1+i\zeta_{(-)}^{\pm})
\Gamma(i\zeta_{(-)}^{\mp})}.
\end{eqnarray}
We can see that the Bogoliubov coefficients are independent of $\rho$, since they are functions of only the dimensionless mass and dimensionless wave number. The  Dirac  field undergoes a
$\epsilon$-dependent Bogoliubov transformation, where
$\epsilon$ is the parameter to be estimated.

To find the relation between the asymptotic past and future vacuum states, we use the relationship between the operators
$b_{in}(\tilde{k})=\left[{{\cal A}^{-}_{\tilde{k}}}^{\ast}b_{out}(\tilde{k})-{{\cal B}^{-}_{\tilde{k}}}^{\ast}%
\chi(\tilde{k})a_{out}^{\dagger}(-\tilde{k})\right]$, where $\chi(\tilde{k})=\frac{\tilde{\mu}_{out}}{\mid \tilde{k}\mid}(1-\frac{\tilde{\omega}_{out}}{\tilde{\mu}_{out}})$ with $\mu_{out}=\rho\tilde{m}\sqrt{a^2(\eta\rightarrow +\infty)}$. The annihilation  and  creation
operators of the Dirac fields satisfy the canonical anticommutation
$\{b_{out}(\tilde{k}),b_{out}^{\dagger}(\tilde{k}')\}=\delta_{\tilde{k}\tilde{k}'}$ and  $\{b_{out}(\tilde{k}),b_{out}(\tilde{k}')\}=\{b_{out}^{\dagger}(\tilde{k}),b_{out}^{\dagger}(\tilde{k}')\}=0$, where $\{.,.\}$ denotes the anticommutator. The creation operator $a_{out}^{\dagger}(-\tilde{k})$, acting upon the vacuum state $||0\rangle\rangle_{out}$ \cite{antifermi}, will populate the asymptotic future vacuum with a single antifermion. Here we denote the states in the antisymmetric fermionic Fock space by double-lined Dirac notation $||.\rangle\rangle$ \cite{antifermi} rather than  $|.\rangle$. Imposing $b_{in}(\tilde{k})||0\rangle\rangle_{in}=0$ and the vacuum normalization $_{in}\langle\langle 0||0\rangle\rangle_{in}=1$,
 we obtain the asymptotically past vacuum state in terms of the
 asymptotic future fermionic Fock basis
 \begin{equation}
||0\rangle\rangle_{in}=\bigotimes_k\frac{||0\rangle\rangle_{out}-\gamma^{F\ast}_{\tilde{k}}\chi_{\tilde{k}}||1_k
1_{-k}\rangle\rangle_{out} }{\sqrt{1+|\gamma_{\tilde{k}}^{F}\chi_{\tilde{k}}|^{2}}},
 \label{vacin}
\end{equation}
where
\begin{eqnarray}
\left|\gamma^{F}_{\tilde{k}}\right|^2 =\frac{\zeta_{(-)}^{+}
\zeta_{(+)}^{+}}{\zeta_{(-)}^{-}\zeta_{(+)}^{-}}
\ \!\!\frac{\sinh\left[\pi \zeta_{(-)}^{-}\right]
\sinh\left[\pi\zeta_{(-)}^{+}\right]}{\sinh
\left[\pi\zeta_{(+)}^{+}\right]\sinh
\left[\pi\zeta_{(+)}^{-}\right]}.\label{gamma-F}
\end{eqnarray}
We can see that a single qubit prepared initially in a unexcited state $||0\rangle\rangle_{in}$  evolves to an entangled state in the particle number degree of freedom in the expanding universe.  The system is entangled by gravity during the expanding and the information about the volume ratio $\epsilon$ is codified in the final state.

\section{Fisher information and quantum metrology }
The aim of
quantum metrology is to determine the value of a parameter with
higher precision than classical approaches, by using entanglement
and other quantum resources  \cite{advances}. A key concept for
quantum metrology is the quantum Fisher information \cite{paris},
which relates with a measurement outcome $\xi$ of a positive operator
valued measurement (POVM) $\{\hat{E}(\xi)\}$ and takes the form of
\begin{eqnarray}
\mathcal{F}_{\xi}(\epsilon)=\sum_{\xi}P(\xi|\epsilon)
\left(\frac{\partial\ln{P}(\xi|\epsilon)}{\partial\epsilon}\right)^{2},
\end{eqnarray}
where $P(\xi|\epsilon)$ is the probability with respect to a
chosen POVM and the corresponding measurement result $\xi$,
and $\epsilon$ is the parameter to be estimated.  According to
the classical Cram\'{e}r-Rao inequality \cite{Cramer:Methods1946},
the mean square fluctuation of the unbiased error for $\epsilon$ is
\begin{eqnarray}
\Delta\epsilon\geq1/\sqrt{\mathcal{F}_{\xi}(\epsilon)}.
\end{eqnarray}
Furthermore, the Fisher information is
$N\mathcal{F}_{\xi}(\epsilon)$ for $N$ repeated trials of
measurement, which leads to Cram\'{e}r-Rao Bound(CRB) on
the phase uncertainty, $N(\Delta\epsilon)^2\geq1/\mathcal{F}_{\xi}
(\epsilon) $ which gives the maximum
precision in $\epsilon$ for a \emph{particular} measurement scheme.
Optimizing over all the possible quantum measurements
provides an lower bound~\cite{Braunstein1994}
\begin{equation}\label{Cramer-Rao}
N(\Delta \epsilon)^{2}\geq\frac{1}{\mathcal{F}_{\xi}(\epsilon)}\geq \frac{1}{\mathcal{F}_{Q}(\epsilon)},
\end{equation}
where $\mathcal{F}_{Q}(\epsilon)$ is the \emph{quantum}
Fisher information. Usually, the QFI can be calculated
from the density matrix of the state by the
maximum $\mathcal{F}_{Q}(\epsilon):=\max_{\{ \hat{E}(\xi)\}
}{\mathcal{F}}_{\xi}(\epsilon)$, which is saturated by a particular POVM and
defined in terms of the symmetric logarithmic derivative
(SLD) Hermitian operator $\mathcal{L}_\epsilon$
\begin{equation}
\mathcal{F}_{Q}(\epsilon)={\rm Tr}\left(\rho_\epsilon,\mathcal{L}_\epsilon^{2}\right)
  ={\rm Tr}\left[\left(\partial_{\epsilon}\,\rho_{\epsilon}\right)\,
  \mathcal{L}_\epsilon\right],
  \label{eq:QFI1}
\end{equation}
where
\begin{equation}
  \partial_{\epsilon}\,\rho_{\epsilon}=\frac{1}{2}\left\{\rho_\epsilon,\mathcal{L}_\epsilon\right\},
  \label{eq:SLD}
\end{equation}
where $\partial_{\epsilon}\equiv\frac{\partial}{\partial\epsilon}$
and $\left\{\,\cdot\,,\,\cdot\,\right\}$
denotes the anticommutator. Although now we have a concise definition of
QFI, the calculation of $\mathcal{L}_{\epsilon}$ is somewhat not easy. Thus, we should rephrase the QFI by a spectrum decomposition of the state as $\rho_{\epsilon}=\sum_{i=1}^N \lambda_i|\psi_i\rangle\langle\psi_i|$, which has the form of
\begin{equation}
\mathcal{F}_{Q}(\epsilon)=2\sum_{m,n}^N\frac{|\langle\psi_m|
\partial_{\epsilon}\rho_{\epsilon}|\psi_n\rangle|^2}{\lambda_m+\lambda_n},
\end{equation}
where the eigenvalues $\lambda_i\geq0$ and $\sum_i^N\lambda_i=1$.
For a quantum state with non-full-rank density matrix, the QFI can be
expressed as \cite{Zhong}
\begin{eqnarray}
\mathcal{F}_{Q}(\epsilon)=
  \sum_{m'}\frac{\left(\partial_{\epsilon}\lambda_{m'}\right)^{2}}{\lambda_{m'}}+
2\sum_{m\neq n}\frac{\left(\lambda_{m}-\lambda_{n}\right)^{2}}{\lambda_{m}+\lambda_{n}}
  \left|\left\langle\psi_{m}|\partial_{\epsilon}\psi_{n}\right\rangle \right|^{2},
  \label{mixed}
\end{eqnarray}
where the summations involve sums over all $\lambda_{m'}\neq0$ and
$\lambda_{m}+\lambda_{n}\neq0$, respectively.

\section{QFI and optimal bounds for parameter estimations in RW universe}
Our aim is to study how precisely one can in principle estimate
the cosmological parameters that appear in the expansion of universe.
Specifically, we will estimate the volume ratio $\epsilon$  of the universe based on the
quantum Cram$\mathrm{\acute{e}}$r-Rao inequality \cite{Cramer:Methods1946}. Thus we should at first calculate the operational QFI in terms of the Bogoliubov coefficients which encode in the estimated parameter. The reduced density matrix  corresponding to particle states of fixed momentum $\tilde{k}$ is found to be
\begin{equation}
\varrho^F_{\tilde{k}}=\frac{1}{(1+|\gamma_{\tilde{k}}^{F}\chi_{\tilde{k}}|^{2})}(||0\rangle\rangle
\langle\langle0||+|\gamma_{\tilde{k}}^{F\ast}\chi_{\tilde{k}}|^{2}||1_k
\rangle\rangle\langle\langle1_k||),\label{pof}
\end{equation}
which is obtained after
tracing out the antiparticle modes $n_{-k}$ and all other modes with
different $k$ since all modes decouple during the evolution of spacetime except for modes with opposite momenta $k$ \cite{fredericivy1}. Assuming that now we live in the universe that corresponds to
the asymptotic future of the spacetime, we can  estimate the
 volume ratio of the expanding universe by taking measurements
on the particle state Eq. (\ref{pof}).

Generally, an estimation scheme can be performed in three
steps: probe preparations, evolution of the probes in the
dynamic system, and measure the probes after the evolution \cite{advances}.
In this system, the state between the created fermions and anti-fermions act as the probe. For different positive operator valued measurements (POVM) $\hat{E}(\xi)$, different values of Fisher information can be obtained.
The aim of this paper is looking for highest precision for the estimation. Then we should find  the optimal estimation scheme, i.e. finding the
optimal POVM that allows us to get the largest Fisher information, i.e., the QFI. Now our main task is to find the optimal estimation strategy, i.e. finding an optimal measurement obtaining the QFI. In this paper the eigenvector $|\psi_m\rangle$ of the final state Eq.(7) is easily to be calculated. Then the projective measurement can be constituted by the eigenvectors $|\psi_m\rangle$ of the final state and the measured probabilities $p(\xi\mid T)$ are exactly the eigenvalues  $p_m$ of the final state.  The symmetric logarithmic derivative
operator $\mathcal{L}_{\epsilon}$ for the
parameter $\epsilon$ is
$\mathcal{L}_{\epsilon}=2\langle\langle
j||\partial_{\epsilon}(\varrho^F_{\tilde{k}})
||l\rangle\rangle$,
where $j(\neq l)=(0,1)$ and the QFI is found to be
\begin{eqnarray}
 \mathcal{F}_{Q}(\epsilon)=&(1+|\gamma_{\tilde{k}}^{F}\chi_{\tilde{k}}|^{2})
[\partial_{\epsilon}(\frac{1}{1+|\gamma_{\tilde{k}}^{F}\chi_{\tilde{k}}|^{2}})]^2
+\frac{1+|\gamma_{\tilde{k}}^{F}\chi_{\tilde{k}}|^{2}}{|\gamma_{\tilde{k}}^{F\ast}\chi_{\tilde{k}}|^{2}}
[\partial_{\epsilon}(\frac{|\gamma_{\tilde{k}}^{F\ast}\chi_{\tilde{k}}|^{2}}
{1+|\gamma_{\tilde{k}}^{F}\chi_{\tilde{k}}|^{2}})]^2,\label{eqpofq}
\end{eqnarray}
from which we can see that the expanding parameter $\epsilon$ is involved in the expression of QFI.  It is worth mentioning that particles are created by the expansion of spacetime even if initially the field was in its vacuum state. Pairs of entangled particles with opposite momentum $\tilde{k}$ are produced. We can exploit these particles for our purposes. In some extreme cases, for example when $\tilde{m}=\tilde{k}=1$ and $\epsilon\rightarrow\infty$, equation (8) becomes $
\gamma^{F}_{\tilde{k}} \rightarrow\frac{1-e^{(1-\sqrt{2})\pi}}{e^{\sqrt{2}\pi+1}-1},
$
and the QFI vanishes because the partial derivatives $\partial_{\epsilon}(\frac{1}{1+|\gamma_{\tilde{k}}^{F}\chi_{\tilde{k}}|^{2}})$ and $\partial_{\epsilon}(\frac{|\gamma_{\tilde{k}}^{F\ast}\chi_{\tilde{k}}|^{2}}
{1+|\gamma_{\tilde{k}}^{F}\chi_{\tilde{k}}|^{2}})$ equal to zero.  In other words, the quantum parameter estimation has the lowest precision when $\epsilon\rightarrow\infty$.

\begin{figure}[ht]
\includegraphics[scale=0.8]{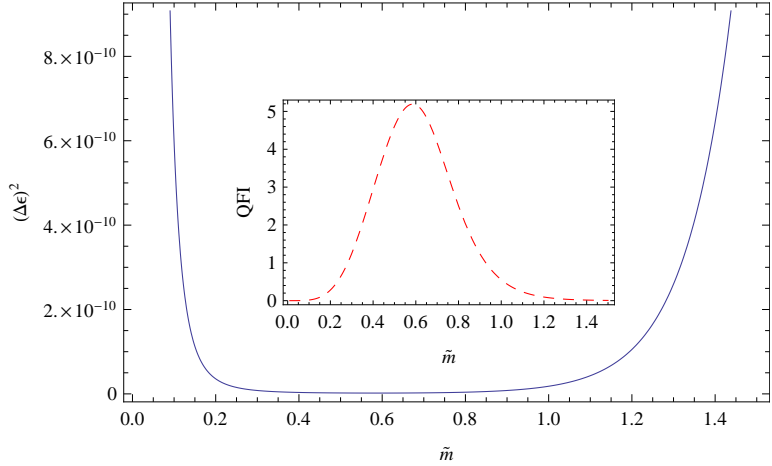}
\caption{(Color online)  The optimal bound  $(\Delta\epsilon)^2$ in the
estimation of the volume ratio $\epsilon$  of the universe over the dimensionless mass $\tilde{m}$ of the Dirac particles. The parameters are fixed with $\epsilon=\tilde{k}=1$. The measurements repeated $n=10^{11}$ times.  The QFI (dashed red line) of the Dirac fields as a function of dimensionless mass $\tilde{m}$ for fixed $\epsilon=\tilde{k}=1$.}\label{Fig1}
\end{figure}

The adjustable parameters in the probe state Eq.(16) are the dimensionless mass $\tilde{m}$ and the dimensionless  momentum $\tilde{k}$ of the particle and antiparticles. Finally, we
proceed to compute the error bounds which are our main quantity of interest. The optimal bound for
the error in the estimation process can be calculated by using Eqs. (\ref{Cramer-Rao}) and (\ref{eqpofq}).
In Fig. (\ref{Fig1}) we plot the optimal bound $(\Delta\epsilon)^2$  in the
estimation of the volume ratio $\epsilon$  as a function of dimensionless mass $\tilde{m}$ of the Dirac fields.
It is found that the bounds
decrease rapidly as the mass for the light Dirac fields increases.
We find that there is a finite range of masses that allows for a better estimation of the volume ratio. Thus one can find a method for obtaining the lowest bound by choosing the final projection
operator to be on the eigenvectors of the particle states with a chosen mass, determined by
the desired outcome. In this way, one can improve the quantum estimation technologies for the volume ratio $\epsilon$ in an expanding universe.

In Fig. (\ref{Fig1}) we also plot the QFI (dashed red line) of the Dirac fields in the expanding
universe as a function of dimensionless mass $\tilde{m}$ for fixed $\epsilon=\tilde{k}=1$. It is interesting to note that
in Fig. 2 of Ref. \cite{fredericivy1}, the behaviour of entanglement for Dirac fields is
very similar with the behaviour of QFI in this
paper. Then we arrive to a conclusion that the precision of the estimation of parameters in
the expanding universe will be improved from the increase
of quantum entanglement. This conclusion corroborate that the precision of quantum parameter estimation
can be enhanced if squeezed or entangled states are used as probe
systems. For some very larger masses, the QFI diminishes due to the disappearance of quantum entanglement.

\begin{figure}[ht]
\includegraphics[scale=0.8]{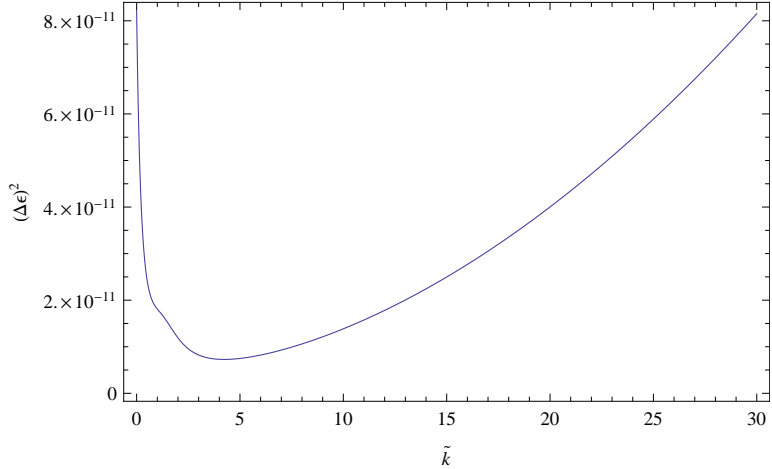}
\caption{(Color online) The optimal bound  $(\Delta\epsilon)^2$ in the
estimation of the volume ratio $\epsilon$  of the universe over $\tilde{k}$. The parameters are fixed with $\epsilon=\tilde{m}=1$. The measurements repeated $n=10^{11}$ times.}\label{Fig3}
\end{figure}

In Fig. (\ref{Fig3}) we plot the optimal bounds
$(\Delta\epsilon)^2$  as a function of $\tilde{k}$. The minimum bound is obtained by numerical optimization over $\tilde{k}$.  It is shown that the bond $(\Delta\epsilon)^2$ deceases at first and then increase as the increasing $\tilde{k}$, which means that the minimum possible bound for the estimation of the volume ratio $\epsilon$
can be find for some optimal $\tilde{k}$. In other words, one can get the highest precision by choosing some fermions with some certain $\tilde{k}$ in the quantum measuring process.

\section{Conclusions}

We studied cosmological quantum metrology by incorporating the effect of universe expansion in quantum parameter estimation. It is shown that the expanding spacetime background changes the value of QFI.  At the same time, information about the history of the expanding universe can be extracted from measurements on quantum states in the asymptotic future region. We show that the optimal precision of the estimations depend sensitively on the dimensionless mass $\tilde{m}$ and dimensionless momentum $\tilde{k}$ of the Dirac particles.  We find that the precision of the estimation of a parameter in
the expanding universe will benefit from the increase
of quantum entanglement, which corroborates the fact that the precision can be
enhanced if squeezed or entangled states are used as probe systems. It was also shown that the precision of the estimation can be improved by choosing the probe state as an energy eigenvector (i.e., defined by particle number). Such result can be understood from the fact that the largest quantum Fisher information can be obtained by projective measurements corresponding to eigenvectors of the particle states of specific masses and momentums.

\begin{acknowledgments}
This work is supported the National Natural Science Foundation
of China under Grant No. 11305058, No. 11175248, No. 11475061, the Doctoral Scientific Fund Project of the Ministry of Education of China under Grant No. 20134306120003, Postdoctoral Science Foundation of China under Grant No. 2014M560129, and the Strategic Priority Research Program of the Chinese Academy of Sciences (under Grant No. XDB01010000).	
\end{acknowledgments}

\end{document}